\documentclass[letterpaper, 10 pt, conference]{ieeeconf}
\IEEEoverridecommandlockouts                              
\usepackage{mathrsfs}
\usepackage{amsfonts}
\usepackage{amsmath,amssymb}
\usepackage{graphicx}
\usepackage{subfigure}
\usepackage{epstopdf}
\usepackage{subfigure}
\usepackage{epic}
\usepackage{tikz}
\usepackage{algorithm}
\usepackage{algpseudocode}
\usepackage{graphics} 
\usepackage{color}
\usepackage[justification=centering]{caption}
\usetikzlibrary{arrows,shapes,chains}
\usepackage{cite}
\usepackage{algorithm}
\usepackage{algpseudocode}

\floatname{algorithm}{Algorithm}

\newtheorem{definition}{Definition}
\newtheorem{theorem}{Theorem}

\newtheorem{remark}{Remark}
\newtheorem{assumption}{Assumption}
\newtheorem{proposition}{Proposition}
\newtheorem{example}{Example}

\begin{document}
%
\title{\LARGE{\bf{Approximately symbolic models for a class of continuous-time nonlinear systems}}}
%
%

\author{Pian Yu and Dimos V. Dimarogonas
\thanks{This work was supported in part by the Swedish Research Council (VR), the European Research Council (ERC), the Swedish Foundation for Strategic Research (SSF) and the Knut and Alice Wallenberg Foundation (KAW).}
\thanks{The authors are with School of Electrical Engineering and Computer Science, KTH Royal Institute of Technology, 10044 Stockholm, Sweden.
        {\tt\small piany@kth.se, dimos@kth.se}}
}

\maketitle

\begin{abstract}
Discrete abstractions have become a standard approach to assist control synthesis under complex specifications. Most techniques for the construction of discrete abstractions are based on sampling of both the state and time spaces, which may not be able to guarantee safety for continuous-time systems. In this work, we aim at addressing this problem by considering only  state-space abstraction. Firstly, we connect the continuous-time concrete system with its discrete (state-space) abstraction with a control interface. Then, a novel stability notion called controlled globally asymptotic/practical stability with respect to a set is proposed. It is shown that every system, under the condition that there exists an admissible control interface such that the augmented system (composed of the concrete system and its abstraction) can be made controlled globally practically stable with respect to the given set, is approximately simulated by its discrete abstraction. The effectiveness of the proposed
results is illustrated by a simulation example.
\end{abstract}
%

\IEEEpeerreviewmaketitle

\section{Introduction}

In recent years, discrete abstractions have become one of the standard approaches for control synthesis in the context of complex dynamical systems and specifications \cite{Tabuada09}. It allows one to leverage computational tools developed for discrete-event systems \cite{Ramadge87,Kumar95,Cassandras99} and games on automata \cite{Arnold03,Madhusudan03} to assist control synthesis for specifications difficult to enforce with conventional control design methods. Moreover, if the behaviors of the original (continuous) system (referred to as the concrete system) and the abstract system (with discrete state-space) can be formally related by an inclusion or equivalence relation, the synthesized controller is known to be correct by design \cite{Girard12}.

For a long time, bisimulation relations was a central notion to deal with complexity reduction \cite{Milner89,Park81}. It was later pointed out that requiring strict equality of observed behaviors is often too strong \cite{Alur00}. To this end, a new notion called approximate bisimulation was introduced in \cite{Girard_Pappas07}. Based on the notion of incrementally (input-to-state) stability \cite{Angeli02}, approximately bisimilar symbolic models were built and extended to various systems \cite{Girard_Pola10,Zamani14}. However, incremental (input-to-state) stability is a stronger property than (input-to-state) stability for dynamical control systems, which makes its applicability still restrictive. In \cite{Zamani12}, the authors relax this requirement by only assuming Lipschitz continuous and incremental forward completeness, and an approximately alternating simulation relation is established by over-approximating the behavior of the concrete system. However, as recently pointed out in \cite{Reissig17}, this approach may result in a refinement complexity issue.

This paper investigates the construction of symbolic models for continuous-time nonlinear systems based on the notion of approximate simulation. It improves upon most of the existing results in two aspects: 1) by being applicable to nonlinear systems that are not incrementally stable and 2) by not requiring time-space abstraction. In the first aspect, we propose novel stability notions, called controlled globally asymptotic/practical stability with respect to a given set $\Omega$ (C-$\Omega$-GA/PS). These are properties defined on both the concrete system and the abstract system via an admissible control interface. It is shown that for (concrete) systems which are not incrementally stable, the C-$\Omega$-GA/PS properties can still be satisfied. In the second aspect, we show that the abstract system can be constructed such that the concrete system is $\varepsilon$-approximately simulated by the abstraction without time-space discretization. This point is crucial for safety-critical applications, in which it is necessary that the trajectories of the concrete system and the abstract system are close enough at all time instants. In particular, the application to a class of incremental quadratic nonlinear systems is investigated.

The introduction of the control interface is inspired by the hierarchical control framework \cite{Girard_Pappas09,Fu13,Yang17,Smith18(2)}, in which an interface is built between a high dimensional concrete system and a simplified low dimensional abstraction of it. In this paper, we propose to build a control interface between the continuous concrete system and its discrete (state-space) abstraction. Moreover, the consideration of bounded input set (the input set considered in \cite{Girard_Pappas09,Fu13,Yang17,Smith18(2)} is unbounded) brings additional difficulty to constructing the interface. Therefore, the results in this paper are essentially novel and improved with respect to the existing work.

The remainder of this paper is organized as follows. In Section II, notation and preliminaries are provided. New stability notions are defined in Section III and the main results are presented in Section IV. In Section V, an application to incremental quadratic nonlinear systems is provided. An illustrative example is given in Section VI and Section VII concludes the paper.

\section{Preliminaries}

\subsection{Notation}
Let $\mathbb{R}:=(-\infty, \infty)$, $\mathbb{R}_{\ge 0}:=[0, \infty)$, $\mathbb{R}_{> 0}:=(0, \infty)$, $\mathbb{Z}_{> 0}:=\{1,2,\ldots\}$ and $\mathbb{Z}_{\ge 0}:=\{0,1,2,\ldots\}$. Denote $\mathbb{R}^n$ as the $n$ dimensional real vector space, $\mathbb{R}^{n\times m}$ as the $n\times m$ real matrix space. $I_n$ is the identity matrix of order $n$ and $1_n$ is the column vector of order $n$ with all entries equal to one. When there is no ambiguity, we use $0$ to represent a matrix with proper dimensions and all its elements equal to $0$. $[a, b]$ and $[a, b[$ denote closed and right half-open intervals with end points $a$ and $b$. For $x_1\in\mathbb{R}^{n_1}, \ldots, x_m\in\mathbb{R}^{n_m}$, the notation $(x_1, x_2, \ldots, x_m)\in \mathbb{R}^{n_1+n_2+\cdots +n_m}$ stands for $[x_1^T, x_2^T, \ldots, x_m^T]^T$. Let $\left|\lambda\right|$ be the absolute value of a real number $\lambda$, and $\|x\|$ and $\|A\|$ be the Euclidean norm of vector $x$ and matrix $A$, respectively. Given a function $f: \mathbb{R}_{\ge 0}\to \mathbb{R}^n$, the supremum of $f$ is denoted by $\|f\|_\infty$, which is given by $\|f\|_\infty:=\sup\{\|f(t)\|, t\ge 0\}$ and $\|f\|_{[0, \tau)}:=\sup\{\|f(t)\|, t\in [0, \tau)\}$. A function $f$ is called bounded if $\|f\|_\infty<\infty$. Given a set $S$, the boundary of $S$ is denoted by $F_r(S)$. Given two sets $S_1, S_2$, the notation $S_1\setminus S_2:=\{x| x\in S_1 \; \wedge \; x\notin S_2\}$ stands for the set difference, where $\wedge$ represents the logic operator AND.

A continuous function $\gamma: \mathbb{R}_{\ge 0}\to \mathbb{R}_{\ge 0}$ is said to belong to class $\mathcal{K}$ if it is strictly increasing and $\gamma(0)=0$; $\gamma$ is said to belong to class $\mathcal{K}_{\infty}$ if $\gamma\in \mathcal{K}$ and $\gamma(r)\to \infty$ as $r\to \infty$. A continuous function $\beta: \mathbb{R}_{\ge 0}\times \mathbb{R}_{\ge 0}\to \mathbb{R}_{\ge 0}$ is said to belong to class $\mathcal{K}\mathcal{L}$ if for each fixed $s$, the map $\beta(r, s)$ belongs to class $\mathcal{K}_{\infty}$ with respect to $r$ and, for each fixed $r$, the map $\beta(r, s)$ is decreasing with respect to $s$ and $\beta(r, s)\to 0$ as $s\to \infty$. For a set $\mathcal{A}\subseteq \mathbb{R}^n$ and any $x\in \mathbb{R}^n$, we denote by, $\rm{d}(x, \mathcal{A})$, the point-to-set distance, defined as $\rm{d}(x, \mathcal{A})=\inf_{y\in \mathcal{A}}\{\|x-y\|\}.$

\subsection{System properties}

Consider a dynamical system of the form
\begin{equation}\label{cs}
  \Sigma:\left\{\begin{aligned}
  \dot x_1(t)&= f(x_1(t), u(t))\\
  y_1(t)&=h(x_1(t)),
  \end{aligned}\right.
\end{equation}
where $x_1(t)\in \mathbb{R}^{n}, y_1(t)\in \mathbb{R}^{l}, u(t)\in U\subseteq \mathbb{R}^{m}$ are the state, output and control input of the system, respectively. We assume that $f: \mathbb{R}^{n}\times U \to \mathbb{R}^{n}$ is a continuous map and the vector field $f$ is such that for any input in $U$, any initial condition in $\mathbb{R}^{n}$, this differential equation has a unique solution. Throughout the paper, we will refer to $\Sigma$ as the concrete system. Let ${\bf U}_\tau=\{u: [0, \tau[ \to U\}$ be a set of continuous functions of time from intervals of the form $[0, \tau[$ to $U$ with $\tau>0$, then we define $\mathcal{U}=\cup_{\tau>0}\bf{U}_\tau$. In addition, we use $\text{dom}(u)$ to represent the domain of function $u$.

A curve $\xi: [0, \tau[ \to \mathbb{R}^n$ is said to be a trajectory of $\Sigma$ if there exists input $u\in {\bf U}_\tau$ satisfying $\dot{\xi}(t)=f(\xi(t), u(t))$ for almost all $t\in [0, \tau[$. A curve $\zeta: [0, \tau[ \to \mathbb{R}^l$ is said to be an output trajectory of $\Sigma$ if $\zeta(t)=h(\xi(t))$ for almost all $t\in [0, \tau[$, where $\xi$ is a trajectory of $\Sigma$. We use $\xi(\xi_0, u, t)$ to denote the trajectory point reached at time $t$ under the input signal $u\in {\bf U}_t$ from initial condition $\xi_0$.

\begin{definition}\cite{Angeli99}
A system is called forward complete (FC) if for every
initial condition $x_0\in \mathbb{R}^{n}$ and every input signal $u\in \mathcal{U}$, the
solution is defined for all $t\ge 0$.
\end{definition}

\begin{definition}[Definition 4.13 \cite{Goedel12}]
Given $\varepsilon>0$, two output trajectories $\zeta_1: [0, \tau[\to \mathbb{R}^l$ and $\zeta_2: [0, \tau[\to \mathbb{R}^l$ are $\varepsilon$-close if $\|\zeta_1(t)-\zeta_2(t)\|\le \varepsilon, \forall t\in [0, \tau[.$
\end{definition}

\section{Controlled globally asymptotic or practical stability with respect to a set}

In this paper, the abstraction technique developed in \cite{Girard_Pola10} is applied, where the state-space is approximated by the lattice
\begin{equation}\label{sabs0}
  [\mathbb{R}^n]_{\eta}=\Big\{q\in \mathbb{R}^n| q_i=k_i\frac{2\eta}{\sqrt{n}}, k_i\in \mathbb{Z}, i=1, \ldots, n\Big\},
\end{equation}
where $\eta\in \mathbb{R}_{\ge 0}$ is a state-space discretization parameter. Define the associated quantizer $Q_\eta: \mathbb{R}^n\to [\mathbb{R}^n]_{\eta}$ as $Q_\eta(x)=q$ if and only if $|x_i-q_i|\le \eta/\sqrt{n}, \forall i=1, \ldots n$. Then, one has $\|x-Q_\eta(x)\|\le \eta, \forall x\in \mathbb{R}^n$.

The abstract system is obtained by applying the state abstraction (\ref{sabs0}), which is given by
\begin{equation}\label{abs}
  \Sigma':\left\{\begin{aligned}
  x_2(t)&= Q_\eta(\xi(Q_\eta(x_1(0)), v, t))\\
  y_2(t)&=h(x_2(t)),
  \end{aligned}\right.
\end{equation}
where $x_2(t)\in [\mathbb{R}^n]_\eta$. In addition, $v(t)\in U'$ is the control input for the abstract system. We note that the set $U'$ is a design parameter, which will be specified later. Let $\mathcal{U}'=\cup_{\tau>0}\bf{U}'_\tau$, where ${\bf {U}}'_\tau=\{u: [0, \tau[ \to U'\}$. The trajectory and output trajectory of $\Sigma'$ are denoted by $\xi'$ and $\zeta'$, respectively.

The input $u(t)$ of the concrete system (\ref{cs}) will be synthesized hierarchically via the abstract system (\ref{abs}) with a control interface $u_v: U'\times \mathbb{R}^n\times [\mathbb{R}^n]_\eta \to U$, which is given by
\begin{equation*}
  u(t)=u_v(v(t), x_1(t), x_2(t)).
\end{equation*}
Then, the augmented control system $\hat\Sigma$ is defined as
\begin{equation}\label{hatS}
  \hat\Sigma: \left\{\begin{aligned}
  \dot x_1(t)&=f(x_1(t), u_v(v(t), x_1(t), x_2(t)))\\
  x_2(t)&= Q_\eta(\xi(Q_\eta(x_1(0)), v, t))\\  y_1(t)&=h(x_1(t))\\
  y_2(t)&=h(x_2(t))
  \end{aligned}\right.
\end{equation}
and we denote $x(t)=(x_1(t), x_2(t))\in \hat X:=\{(z, z'): z\in \mathbb{R}^n, z'\in [\mathbb{R}^n]_\eta\}, y(t)=(y_1(t), y_2(t))\in \hat Y:=\{(z, z'): z\in \mathbb{R}^l, z'\in \mathbb{R}^l\}$ and $\hat u(t)=(u_v(t), v(t))\in \hat U:=\{(z, z'): z\in U, z'\in U'\}$. Note that $x_2(0)=Q_\eta(x_1(0))$, and then one has $\|x_1(0)-x_2(0)\|=\|x_1(0)-Q_\eta(x_1(0))\|\le \eta, \forall x_1(0)\in \mathbb{R}^n$. Therefore, one can define
\begin{equation}\label{hatX0}
  \hat X_0:=\{(x_1, x_2)\in \hat X| \|x_1-x_2\|\le \eta\}
\end{equation}
as the set of initial states for $\hat\Sigma$. To guarantee that $u(t)=u_v(v(t), x_1(t), x_2(t))\in U, \forall t$, we propose the following definition.

\begin{definition}\label{def3}
The control interface $u_v: U'\times \mathbb{R}^n\times [\mathbb{R}^n]_\eta \to U$ is called \emph{admissible} to $\hat\Sigma$ if $u_v(v(t), \xi(\xi_0, u_v, t), \xi'(\xi'_0, v, t))\in U$, $\forall (\xi_0, \xi'_0)\in \hat X_0, \forall v(t)\in U'$.
\end{definition}

Let ${\bf {\hat U}}_\tau=\{\hat u: [0, \tau[ \to \hat U\}$ and $\hat{\mathcal{U}}=\cup_{\tau>0}\bf{\hat U}_\tau$. The trajectory of $\hat\Sigma$ will be denoted by $\hat\xi$. The diagonal set $\Omega\subseteq \mathbb{R}^{2n}$ is defined as:
\begin{equation}\label{Omega}
  \Omega=\{z\in \mathbb{R}^{2n}| \exists x\in \mathbb{R}^{n}: z=(x, x)\}.
\end{equation}
Then, we introduce the following definitions which are inspired by \cite{Zamami13}.

\begin{definition}\label{def4}
The augmented control system $\hat\Sigma$ is called \emph{controlled globally asymptotically stable with respect to the set $\Omega$} (C-$\Omega$-GAS) if it is FC and there exists an admissible control interface $u_v: U'\times \mathbb{R}^n\times [\mathbb{R}^n]_\eta \to U$, and a $\mathcal{K}\mathcal{L}$ function $\beta$ such that for any $t\in \mathbb{R}_{\ge 0}$ and any $(x_0, x'_0)\in \hat X_0$, the following condition is satisfied:
\begin{equation*}
 {\rm{d}}(\hat\xi(\hat\xi_0, \hat u, t), \Omega)\le \beta({\rm{d}}(\hat\xi_0, \Omega), t),
\end{equation*}
where $\hat\xi_0=(x_0, x'_0)$ and $\hat u=(u_v, v)$.
\end{definition}

\begin{definition}\label{def5}
The augmented control system $\hat\Sigma$ is called \emph{controlled globally practically stable with respect to the set $\Omega$} (C-$\Omega$-GPS) if it is FC and there exists an admissible control interface $u_v: U'\times \mathbb{R}^n\times [\mathbb{R}^n]_\eta \to U$, a $\mathcal{K}\mathcal{L}$ function $\beta$ and a bounded function $\omega(t)$, such that for any $t\in \mathbb{R}_{\ge 0}$ and any $(x_0, x'_0)\in \hat X_0$, the following condition is satisfied:
\begin{equation*}
 {\rm{d}}(\hat\xi(\hat\xi_0, \hat u, t), \Omega)\le \beta({\rm{d}}(\hat\xi_0, \Omega), t)+\|\omega\|_\infty.
\end{equation*}
\end{definition}

Moreover, $u_v$ is called an \emph{interface} for $\hat\Sigma$, associated to the C-$\Omega$-GAS (C-$\Omega$-GPS) property.

\begin{remark}
According to Definition \ref{def3}, a general idea on determining the set $U'$ can be provided as follows: firstly, we ignore the input constraint by assuming that $U=\mathbb{R}^m$, finding a (set of) control interface(s) $u_v$ such that $\hat\Sigma$ is C-$\Omega$-GAS (C-$\Omega$-GPS). Then, consider the real input set $U$ and choose $U'$ to be the maximal subset of $U$ such that the control interface $u_v$ is admissible.
\end{remark}

\begin{remark}
We note that C-$\Omega$-GAS and C-$\Omega$-GPS are properties defined on the augmented control system $\hat\Sigma$ rather than the concrete system $\Sigma$. Moreover, for concrete systems that are not incrementally stable, these properties can still hold.
\end{remark}

\begin{example}
Consider the following nonlinear system
\begin{equation}\label{csex2}
  \Sigma:\left\{\begin{aligned}
  \dot x_1(t)&= Ax_1(t)+m\sin(x_1(t))+u(t)\\
  y_1(t)&=x_1(t),
  \end{aligned}\right.
\end{equation}
where $x_1, y_1, u\in \mathbb{R}^n$, $A\in \mathbb{R}^{n\times n}$ is a constant matrix and $m\in \mathbb{R}$ is a constant. One can verify that (\ref{csex2}) is not incrementally (input-to-state) stable. Applying the state abstraction (\ref{sabs0}), then the abstract system can be written as
\begin{equation}\label{asex2}
  \Sigma':\left\{\begin{aligned}
  x_2(t)&= Q_\eta(\varphi(Q_\eta(x_1(0)), v, t))\\
  y_2(t)&=x_2(t),
  \end{aligned}\right.
\end{equation}
where $\dot \varphi(t)=A\varphi(t)+m\sin(\varphi(t))+v(t)$ and $v(t)\in U'$.

Let matrices $P=P^T\succ 0, R$ and a scalar $\alpha>0$ be the solutions to the following linear matrix inequality (LMI)
\begin{equation}\label{lmi}
\begin{aligned}
\left( \begin{array}{l}
A^TP+PA+2R+2\alpha P\;\;P\\
\quad\quad\quad\quad\quad P\quad\quad\quad\quad\quad 0
\end{array} \right)+\left( \begin{array}{l}
m^2I_n\;\;\;\;0_n\\
\;\;0_n\;\;-I_n
\end{array} \right)
 \le 0.
\end{aligned}
\end{equation}
Then, the control interface $u_v$ is designed as \begin{equation}\label{uex2}
\begin{aligned}
  u_v(v(t), x_1(t), x_2(t))=v(t)+P^{-1}R(x_1(t)-x_2(t)).
\end{aligned}
\end{equation}
One can see that $u_v$ is admissible for all $U'\subseteq \mathbb{R}^n$. Moreover, according to Lemma 3.5 of \cite{Zamami13}, one has that $\forall x_1, x_2$, ${\rm d}(x, \Omega)=\|x_1-x_2\|$, where $x=(x_1, x_2)$ and $\Omega$ is defined in (\ref{Omega}). Then, one can verify that ${\rm d}(x(t), \Omega)\le \beta({\rm d}(\|x(0)\|, \Omega), t)+C,$ where
\begin{equation*}
\begin{aligned}
  &\beta(z, t)=\sqrt{\frac{\lambda_{\max}(P)}{\lambda_{\min}(P)}}e^{-(\alpha-\frac{a}{2})t}z,\\
  &C=\Bigg(\sqrt{\frac{\|R^TP^{-1}R\|}{a(2\alpha-a)\lambda_{\min}(P)}}+1\Bigg){\eta}.
\end{aligned}
\end{equation*}
That is, the augmented system $\hat\Sigma:= (\Sigma, \Sigma')$ is C-$\Omega$-GPS. $\square$
\end{example}

\begin{proposition}
Consider the augmented control system $\hat\Sigma$. If $\hat\Sigma$ is C-$\Omega$-GAS, then one has $\forall t\in \mathbb{R}_{\ge 0}$, $\forall (x_0, x'_0)\in \hat X_0$,
\begin{equation*}
 \|\xi(x_0, u_v, t)-\xi'(x'_0, v, t)\|\le \beta(\|x_0-x'_0\|, t),
\end{equation*}
where $\beta$ is the $\mathcal{K}\mathcal{L}$ function defined in Definition \ref{def4}.
\end{proposition}

%

\begin{proposition}
Consider the augmented control system $\hat\Sigma$. If $\hat\Sigma$ is C-$\Omega$-GPS, then one has $\forall t\in \mathbb{R}_{\ge 0}$, $\forall (x_0, x'_0)\in \hat X_0$,
\begin{equation*}
\begin{aligned}
  \|\xi(x_0, u_v, t)&-\xi'(x'_0, v, t)\|\\
 &\le \beta(\|x_0-x'_0\|, t)
 +\|\omega\|_\infty,
 \end{aligned}
\end{equation*}
where $\beta$ is the $\mathcal{K}\mathcal{L}$ function defined in Definition \ref{def5}.
\end{proposition}

The following definition of C-$\Omega$-GAS (C-$\Omega$-GPS) Lyapunov function is motivated by \cite{Girard_Pappas09}.

\begin{definition}\label{def6}
A function $V: \mathbb{R}^n\times \mathbb{R}^n\to \mathbb{R}_{\ge 0}$ is an C-$\Omega$-GAS Lyapunov function and $u_v: U'\times \mathbb{R}^n\times [\mathbb{R}^n]_\eta\to U$ is an associated admissible control interface if there exist $\mathcal{K}_\infty$ functions $\underline{\alpha}, \bar{\alpha}$ such that

i) $\forall x, x'\in \mathbb{R}^n$,
\begin{equation}\label{lf1}
  \underline{\alpha}(\|x(t)-x'(t)\|)\le V(x(t), x'(t))\le \bar{\alpha}(\|x(t)-x'(t)\|);
\end{equation}

ii) $\forall (x_0, x'_0)\in \hat X_0$ and $\forall v(t)\in U'$,
\begin{equation}\label{lf2}
  \frac{\partial V}{\partial x} f(x(t), u_v(v(t), x(t), Q_\eta(x'(t))))+\frac{\partial V}{\partial x'} f(x'(t), v(t))\le 0,
\end{equation}
where $x(t)=\xi(x_0, u_v, t), x'(t)=\xi(x'_0, v, t)$.

Function $V$ is called a C-$\Omega$-GPS Lyapunov function and $u_v: U'\times \mathbb{R}^n\times [\mathbb{R}^n]_\eta\to U$ is an associated admissible control interface, if there exist $\mathcal{K}_\infty$ functions $\underline{\alpha}, \bar{\alpha}, \sigma$, and a constant $\gamma>0$ satisfying condition i) and

iii) $\forall (x_0, x'_0)\in \hat X_0$ and $\forall v(t)\in U'$,
\begin{equation}\label{lf3}
\begin{aligned}
  \frac{\partial V}{\partial x} f(x(t), &u_v(v(t), x(t), Q_\eta(x'(t))))+\frac{\partial V}{\partial x'} f(x'(t), v(t))\le \\
  &-\gamma V(x(t), x'(t))+\sigma(\|\omega\|_\infty).
  \end{aligned}
\end{equation}
\end{definition}

Then, we can get the following theorem.

\begin{theorem}
Consider the augmented control system $\hat\Sigma$ and the set $\Omega$. If $\hat\Sigma$ is FC and there exists a C-$\Omega$-GAS or C-$\Omega$-GPS Lyapunov function and $u_v: U'\times \mathbb{R}^n\times [\mathbb{R}^n]_\eta\to U$ an associated admissible control interface, then, $\hat\Sigma$ is C-$\Omega$-GPS and $u_v$ is the \emph{interface} for $\hat\Sigma$, associated to the C-$\Omega$-GPS property.
\end{theorem}

\textbf{Proof:} In the following, we will consider the case where there exists a C-$\Omega$-GAS Lyapunov function. The other case where there exists a C-$\Omega$-GPS Lyapunov function is similar and hence omitted.

Let $\xi(t):=\xi(Q_\eta(x_1(0), v, t)$ for short. Then, we can get $x_2(t)=Q_\eta(\xi(t)), \forall t\ge 0$ and $x_2(0)=\xi(0)=Q_\eta(x_1(0)$. Let $x_1(t)=\xi(x_1(0), u, t)$, where $u(t)=u_v(v(t), x_1(t), x_2(t))$. Since $(x_1(0), \xi(0))\in \hat X_0$ and $u_v$ is an admissible control interface, one has $u_v\in U, \forall v\in U'$. Moreover, there exists a C-$\Omega$-GAS Lyapunov function, then one has (\ref{lf2}) holds and thus $V(x_1(t), \xi(t))\le V(x_1(0), \xi(0))\le \bar\alpha(\|x_1(0)-\xi(0)\|)$. Then, according to (\ref{lf1}), one can further have $\|x_1(t)-\xi(t)\|\le {\underline\alpha}^{-1}(\bar\alpha(\|x_1(0)-\xi(0)\|))$ and
\begin{equation*}
\begin{aligned}
  \|x_1(t)-x_2(t)\|\le & \|x_1(t)-\xi(t)\|+\|\xi(t)-x_2(t)\|\\
  \le & {\underline\alpha}^{-1}(\bar\alpha(\|x_1(0)-x_2(0)\|))+\eta.
\end{aligned}
\end{equation*}
Therefore, $\hat\Sigma$ is C-$\Omega$-GPS and $u_v$ is the \emph{interface} for $\hat\Sigma$, associated to the C-$\Omega$-GPS property.
$\square$

\section{Symbolic models}

\begin{definition}\label{def8}
Given the concrete system $\Sigma$ in (\ref{cs}) and the abstract system $\Sigma'$ in (\ref{abs}), let $\varepsilon>0$ be a given precision. We say that $\Sigma$ is $\varepsilon$-approximately simulated by $\Sigma'$ if:
\begin{itemize}
\item[i)] $\forall x_0\in \mathbb{R}^n, \exists x'_0\in [\mathbb{R}^n]_\eta$ such that $(x_0, x'_0)\in \hat X_0$,
\item[ii)] $\forall (x_0, x'_0)\in \hat X_0$, $\forall v\in \mathcal{U}', \exists u\in \mathcal{U}$ such that
    \begin{equation*}
      \|h(\xi(x_0, u, t))-h(\xi'(x'_0, v, t))\|\le \varepsilon, \forall t\in \text{dom}(v).
    \end{equation*}
\end{itemize}
where $\hat X_0$ is defined in (\ref{hatX0}).
\end{definition}

\begin{remark}\label{rem4}
According to Definition \ref{def8}, if $\Sigma$ is $\varepsilon$-approximately simulated by $\Sigma'$, then for every output trajectory $\zeta'$ in the abstract system $\Sigma'$, there exists an output trajectory $\zeta$ in the concrete system $\Sigma$ such that $\zeta'$ and $\zeta$ are $\varepsilon$-close.
\end{remark}

\begin{assumption}\label{ass1}
The output function $h: \mathbb{R}^n\to \mathbb{R}^l$ is globally Lipschitz continuous with a Lipschitz constant $\rho$. That is, $\|h(x_1)-h(x_2)\|\le \rho\|x_1-x_2\|, \forall x_1, x_2\in \mathbb{R}^n$.
\end{assumption}

Then, we can get the following results.

\begin{theorem}\label{thm1}
Given the concrete system $\Sigma$ in (\ref{cs}) and the abstract system $\Sigma'$ in (\ref{abs}), let $\varepsilon>0$ be a desired precision. Suppose Assumption \ref{ass1} holds. Assume that there exists a C-$\Omega$-GAS Lyapunov function $V$ and let $u_v$ be the associated admissible control interface. If
\begin{equation}\label{eta}
\underline\alpha^{-1}(\bar\alpha(\eta))+\eta<\frac{\varepsilon}{\rho},
\end{equation}
where ${\bar\alpha}, {\underline\alpha}$ and $\rho$ are defined in Definition \ref{def6} and Assumption \ref{ass1}, respectively, then, $\Sigma$ is $\varepsilon$-approximately simulated by $\Sigma'$.
\end{theorem}

\textbf{Proof:} Given $(x_0, x'_0)\in \hat X_0$, since there exists a C-$\Omega$-GAS Lyapunov function $V$ and $u_v$ is the associated admissible control interface, then, $\forall v(t)\in U'$, one has $u_v(v(t), \xi(x_0, u_v, t), \xi'(x'_0, v, t))\in U$. Now, given an input signal $v\in \mathcal{U}'$, one has $u(t)=u_v(v(t), \xi(x_0, u_v, t), \xi'(x'_0, v, t))\in U, \forall t\in \text{dom}(v)$. Thus, $u\in \mathcal{U}$ and $\text{dom}(u)=\text{dom}(v)$.

Let $q(t)=\xi(x'_0, v, t)), t\in \text{dom}(v)$. Then, one has $x_2(t)=\xi'(x'_0, v, t)=Q_\eta(q(t)), \forall t\in \text{dom}(v)$. According to the state abstraction (\ref{sabs0}), one has $\|x_2(t)-q(t)\|\le \eta$. Let also $x_1(t)=\xi(x_0, u_v, t)), t\in \text{dom}(u)$, where $u_v$ is the admissible control interface. To prove item ii) of Definition \ref{def8}, it is sufficient to prove that $\|h(x_1(t))-h(x_2(t)\|\le \varepsilon, \forall t\in \text{dom}(v)$.

Since (\ref{lf2}) of Definition \ref{def6} holds, one has $V(x_1(t), q(t))\le V(x_0, x'_0)\le \bar\alpha(\|x_0-x'_0\|)\le \bar\alpha(\eta).$ Then, $\|x_1(t)-q(t)\|\le \underline\alpha^{-1}(V(x_1(t), q(t)))\le \underline\alpha^{-1}(\bar\alpha(\eta))$. In addition, $\|x_1(t)- x_2(t)\|\le \|x_1(t)-q(t)\|+\|q(t)-x_2(t)\|\le \underline\alpha^{-1}(\bar\alpha(\eta))+\eta\le {\varepsilon}/\rho,$
and thus $\|h(x_1(t))- h(x_2(t))\|\le \rho\|x_1(t)- x_2(t)\|\le \varepsilon$. Item ii) of Definition \ref{def8} thus holds.

By definition of $[\mathbb{R}^n]_\eta$, for all $x_0\in \mathbb{R}^n$, there exists $x'_0\in [\mathbb{R}^n]_\eta$ such that $\|x_0-x'_0\|\le \eta$. Then, from Assumption \ref{ass1}, $\|h(x_0)-h(x'_0)\|\le \rho\|x_0-x'_0\|\le \varepsilon.$ Hence, $(x_0, x'_0)\in \hat X_0$. Item i) of Definition \ref{def8} holds and thus $\Sigma$ is $\varepsilon$-approximately simulated by $\Sigma'$.
$\square$


For the other case where there exists a C-$\Omega$-GPS Lyapunov function, we need the following additional assumption.

\begin{assumption}\label{ass2}
The class $\mathcal{K}_\infty$ function $\underline{\alpha}: \mathbb{R}_{\ge 0}\to \mathbb{R}_{\ge 0}$ satisfies $\underline\alpha^{-1}(a+b)\le \underline\alpha^{-1}(a)+\underline\alpha^{-1}(b), \forall a, b\in \mathbb{R}_{\ge 0}$.
\end{assumption}

We note that Assumption \ref{ass2} is actually a triangular inequality, which is satisfied by various $\mathcal{K}_\infty$ functions, such as the polynomial functions with nonnegative coefficients.

\begin{theorem}\label{thm2}
Given the concrete system $\Sigma$ in (\ref{cs}) and the abstract system $\Sigma'$ in (\ref{abs}). Let $\varepsilon>0$ be a desired precision. Suppose Assumptions \ref{ass1}-\ref{ass2} hold. Assume that there exists a C-$\Omega$-GPS Lyapunov function $V$ and $u_v$ the associated admissible control interface. If furthermore, one has $\sigma (\|g\|_{\infty }) < \gamma\underline\alpha(\varepsilon/\rho)$ and
\begin{equation}\label{thm2c2}
\begin{aligned}
  \underline\alpha^{-1}(\bar\alpha(\eta))+\eta<\frac{\varepsilon}{\rho}-\underline\alpha^{-1}\Big(\frac{\sigma(\|\omega\|_\infty)}{\gamma}\Big),
\end{aligned}
\end{equation}
then, $\Sigma$ is $\varepsilon$-approximately simulated by $\Sigma'$.
\end{theorem}

\begin{remark}
Note that when the input set $U\neq \mathbb{R}^m$, it is in general difficult to construct an admissible control interface $u_v$ since one needs to guarantee that $u_v\neq \emptyset, \forall v\in U'$. The good news is, for a class of incremental quadratic nonlinear systems \cite{Dalto13}, we show in the next section that it is possible to construct an admissible control interface $u_v$, such that $\Sigma$ is $\varepsilon$-approximately simulated by $\Sigma'$, for any input set $U$.
\end{remark}

\section{Applications}

In this section, we consider a class of incremental quadratic nonlinear systems, for which the systematic construction of the admissible control interface is possible. This kind of nonlinear systems are very useful and include many commonly encountered nonlinearities, such as the globally Lipschitz nonlinearity, as special cases.

Consider the nonlinear system described by
\begin{equation}\label{x2}
\Sigma_1: \left\{\begin{aligned}
{\dot x}(t) =& Ax(t) + Bu(t)+Ep(C_qx+D_qp),\\
y(t)=&Cx(t).
\end{aligned}
\right.
\end{equation}
where $x\in \mathbb{R}^n, y\in \mathbb{R}^l$ and $u\in U =\mathbb{R}^m$ are the state, output and control input of the system, respectively, $p: \mathbb{R}^{l_p}\to \mathbb{R}^{l_e}$ represents the knowing continuous nonlinearity of the system, and $A, B, C, E, C_q, D_q$ are constants matrices of appropriate dimensions.

\begin{definition}\cite{Acikmese11}
Given a function $p: \mathbb{R}^{l_p}\to \mathbb{R}^{l_e}$, a symmetric matrix $M\in \mathbb{R}^{(l_p+l_e)\times (l_p+l_e)}$ is called an incremental
multiplier matrix ($\delta$-MM) for $p$ if it satisfies the following
incremental quadratic constraint ($\delta$-QC) for any $q_1, q_2\in \mathbb{R}^{l_p}$:
\begin{equation}\label{qc}
  {\left( \begin{aligned}
{q_2} - {q_1}\quad\\
p({q_2}) - p({q_1})
\end{aligned} \right)^T}M\left( \begin{aligned}
{q_2} - {q_1}\quad\\
p({q_2}) - p({q_1})
\end{aligned} \right) \ge 0.
\end{equation}
\end{definition}

\begin{remark}
The $\delta$-QC condition (\ref{qc}) includes a broad class
of nonlinearities as special cases. For instance, the globally
Lipschitz condition, the sector bounded nonlinearity, and the positive real nonlinearity $p^TSq\ge 0$ for some symmetric, invertible matrix $S$. Some other nonlinearities that can be expressed using the $\delta$-QC were discussed in \cite{Acikmese11,Dalto13}, such as the case when the Jacobian of
$p$ with respect to $q$ is confined in a polytope or a cone.
\end{remark}

\begin{assumption}\label{ass4}
There exist matrices $P=P^T\succ 0, L$ and a scalar $\alpha>0$ such that the following matrix inequality
\begin{equation}\label{lmi2}
\begin{aligned}
\left( \begin{array}{l}
P(A+BL)+(A+BL)^TP+2\alpha P\;\;PE\\
\quad\quad\quad\quad\quad\quad{E^T}P\quad\quad\quad\quad\quad\quad\quad\quad 0
\end{array} \right)\\
 + {\left( \begin{array}{l}
{C_q}\;\;\;D_q\\
0\;\;\;\;\;I
\end{array} \right)^T}M\left( \begin{array}{l}
{C_q}\;\;\;D_q\\
0\;\;\;\;\;I
\end{array} \right) \le 0
\end{aligned}
\end{equation}
is satisfied, where $M=M^T$ is an $\delta$-MM for function $p$.
\end{assumption}
%

The abstract system (obtained by applying the state-space discretization (\ref{sabs0})) is given by
\begin{equation}\label{xx2}
\Sigma'_1: \left\{\begin{aligned}
\xi(t)=&Q_\eta(\hat x(Q_\eta(x(0)), v, t)),\\
\zeta(t)=&C\xi(t).
\end{aligned}
\right.
\end{equation}
where ${\dot {\hat x}}(t) = A{\hat x} + Bv(t)+Ep(C_q {\hat x}+D_qp)$ and $v\in U'$.

The control interface $u_v: U'\times \mathbb{R}^n \times [\mathbb{R}^n]_\eta \to U$ is designed as
\begin{equation}\label{u2}
u_v(v(t), x(t), \xi(t))=v(t)+L(x(t)-\xi(t)),
\end{equation}
where $L$ is the solution of (\ref{lmi2}).
Then, one can verify that $u_v$ is admissible for all $U'\subseteq \mathbb{R}^m$ since $U=\mathbb{R}^m$.

Then, we get the following result.

\begin{theorem}\label{thm3}
Consider the concrete system (\ref{x2}) and the abstract system (\ref{xx2}). The input $u(t)$ of (\ref{x2}) is synthesized by the control interface (\ref{u2}). Suppose that Assumption \ref{ass4} holds, and the state-space discretization parameter satisfies
\begin{equation*}
\begin{aligned}
\eta\le \frac{\varepsilon}{\|C\|}\frac{\sqrt {ak{\lambda _{\min }}(P)} }{\sqrt {ak{\lambda _{\min }}(P)}+\sqrt {ak{\lambda _{\max }}(P)+\|\hat L\|}},
\end{aligned}
\end{equation*}
where $k=2\alpha-a, 0<a<2\alpha$, $\hat L=L^TB^TPBL$ and $P$ is the solution to Assumption \ref{ass4}. Then, $\Sigma_1$ is $\varepsilon$-approximately simulated by $\Sigma'_1$.
\end{theorem}

\textbf{Proof:} Let $\hat x(t)=\hat x(Q_\eta(x(0)), v, t)$ and $e(t)=\xi(t)-\hat x(t)$, then one has $\|e(t)\|\le \eta, \forall t$. Define $\delta(t)=x(t)-\hat x(t)$. Then, from (\ref{x2}) and (\ref{xx2}) one has
\begin{equation*}
\begin{aligned}
\dot\delta(t)=&A\delta(t)+BL(\delta(t)+e(t))\\
&+E(p(C_q x+D_q p)-p(C_q\hat x+D_q p))\\
=&A_c\delta(t)+BLe(t)+E\Phi_p(x, \hat x),
\end{aligned}
\end{equation*}
where $A_c=A+BL$ and $\Phi_p(x, \hat x)=p(C_qx+D_q p)-p(C_q\hat x+D_q p)$. Post and pre multiplying both sides of inequality (\ref{lmi}) by $(\delta(t), \Phi_p(x, \hat x))$ and its transpose and using condition (\ref{qc}) we obtain $\delta^T P \dot{\delta}\le -\alpha \delta^T P\delta+\delta^T PBL e.$

Consider the following Lyapunov function candidate
\begin{equation}\label{VV}
V(x, \hat x)=(x-\hat x)^T P (x-\hat x).
\end{equation}
Then, one has $\lambda_{\rm{min}}(P)\|x-\hat x\|^2\le V(x, \hat x)\le \lambda_{\rm{max}}(P)\|x-\hat x\|^2$. Taking the derivative of $V$ on $t$, one has
\begin{equation*}
\begin{aligned}
\dot V(x, \hat x)&=2\delta^T P \dot{\delta}\le -kV(x, \hat x)+\frac{1}{a}\|\hat L\|\eta^2.
\end{aligned}
\end{equation*}
Therefore, $V(x, \hat x)$ is a valid C-$\Omega$-GPS Lyapunov function for $\hat\Sigma_1:=(\Sigma_1, \Sigma'_1)$, where $\underline\alpha(x)=\lambda_{\rm min}(P)x^2, \bar\alpha(x)=\lambda_{\rm max}(P)x^2$ and $\sigma (\|\omega\|_{\infty })=({1}/{a})\|\hat L\|\eta^2$. In addition, one can verify that Assumption \ref{ass1} holds with $\rho=\|C\|$, and Assumption \ref{ass2} holds. Then, the conclusion follows by applying Theorem \ref{thm2}.
$\square$

When the input set $U=\mathbb{R}^m$, the construction of admissible control interface $u_v$ is relatively easy since $u_v$ is always admissible. However, in practical applications, input saturations are common constraints.

In the following, we will show that the results in Theorem \ref{thm3} still hold when the set $U\neq \mathbb{R}^m$ with the following modifications. Now, we consider the same control interface $u_v$ as in (\ref{u2}). To guarantee that (\ref{u2}) is admissible, we need to find a set $U'$ such that $u_v\in U, \forall v\in U', \forall (x(0), \xi(0))\in \hat X_0$. From Theorem \ref{thm3}, we have that $\dot V(x(t), \hat x(t))\le -kV(x(t), \hat x(t))+\frac{1}{a}\|\hat L\|\eta^2.$
From the comparison principle, we get that
\begin{equation*}
\begin{aligned}
 V(x(t), \hat x(t))\le &e^{-kt}V(x(0), \hat x(0))+\frac{\|\hat L\|\eta^2}{ak}(1-e^{-kt})\\
 \le &\lambda_{\max}(P)\eta^2+\frac{\|\hat L\|\eta^2}{ak}.
\end{aligned}
\end{equation*}
Then, one can further have $\|x(t)-\hat x(t)\|\le \sqrt{{V(x(t), \xi(t))}/{\lambda_{\min}(P)}}
\le K_1\eta,$ where $K_1=\sqrt{{\lambda_{\max}(P)}/{\lambda_{\min}(P)}+{\|\hat L\|}/{(ak\lambda_{\min}(P))}}$, and $\|x(t)-\xi(t)\|\le \|x(t)-\hat x(t)\|+\|\hat x(t)-\xi(t)\|\le(K_1+1)\eta.$

Define $e_u(t)=u(t)-v(t)$. Then, one has $\|e_u(t)\|=\|L(x(t)-\xi(t))\|
\le \|L\|(K_1+1)\eta$. That is, $\|e_u(t)\|$ is upper bounded and the radius of the upper bound is determined by $\eta$ (due to the special form of control interface that was designed in (\ref{u2})). Let $\tilde U=\big\{z\in U| {\rm d}(z, F_r(U))<\|L\|(K_1+1)\eta\big\},$
be the set of points in $U$, whose distance to the boundary of $U$ is less than $\|L\|(K_1+1)\eta$. Then, by choosing $U'=U\setminus\tilde U$, one can guarantee that $u_v\in U, \forall v\in U'$. Note that one can always find $U'\neq \emptyset$ by letting $\eta$ be small enough since $U'\to U$ when $\eta\to 0$.

\section{Simulation}

Consider the nonlinear system $\Sigma$ given in (\ref{csex2}), where $A=(0.15,0; 0,0.5)$, $m=2$ and $n=2$. The abstract system $\Sigma'$ is given in (\ref{asex2}). Let $\varepsilon=0.5$ be the desired precision. The control interface is given by (\ref{uex2}), where $P=I_2, R=-5I_2$ is the solution of the LMI (\ref{lmi}) by letting $\alpha=2.4$. According to Theorem \ref{thm3}, the desired precision $\varepsilon=0.5$ can be achieved by choosing the state-space discretization parameter $\eta=0.15$.

\begin{figure}[h!]
\centering
\includegraphics[height=4cm,width=9cm]{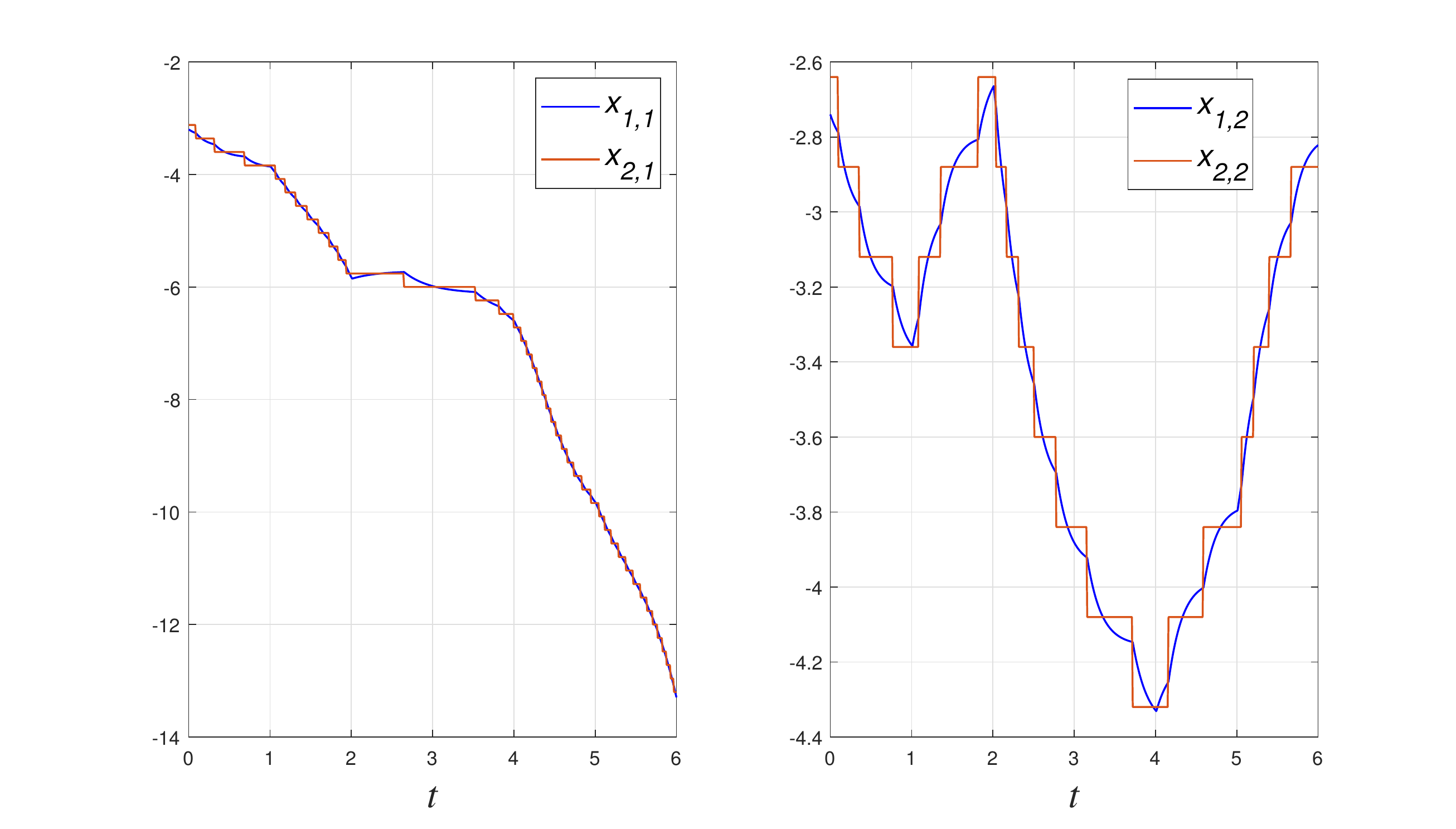}
\caption{The trajectories of the concrete system $\Sigma$ and the abstract system $\Sigma'$. }\label{fig1}
\end{figure}

\begin{figure}[h!]
\centering
\includegraphics[height=4cm,width=9cm]{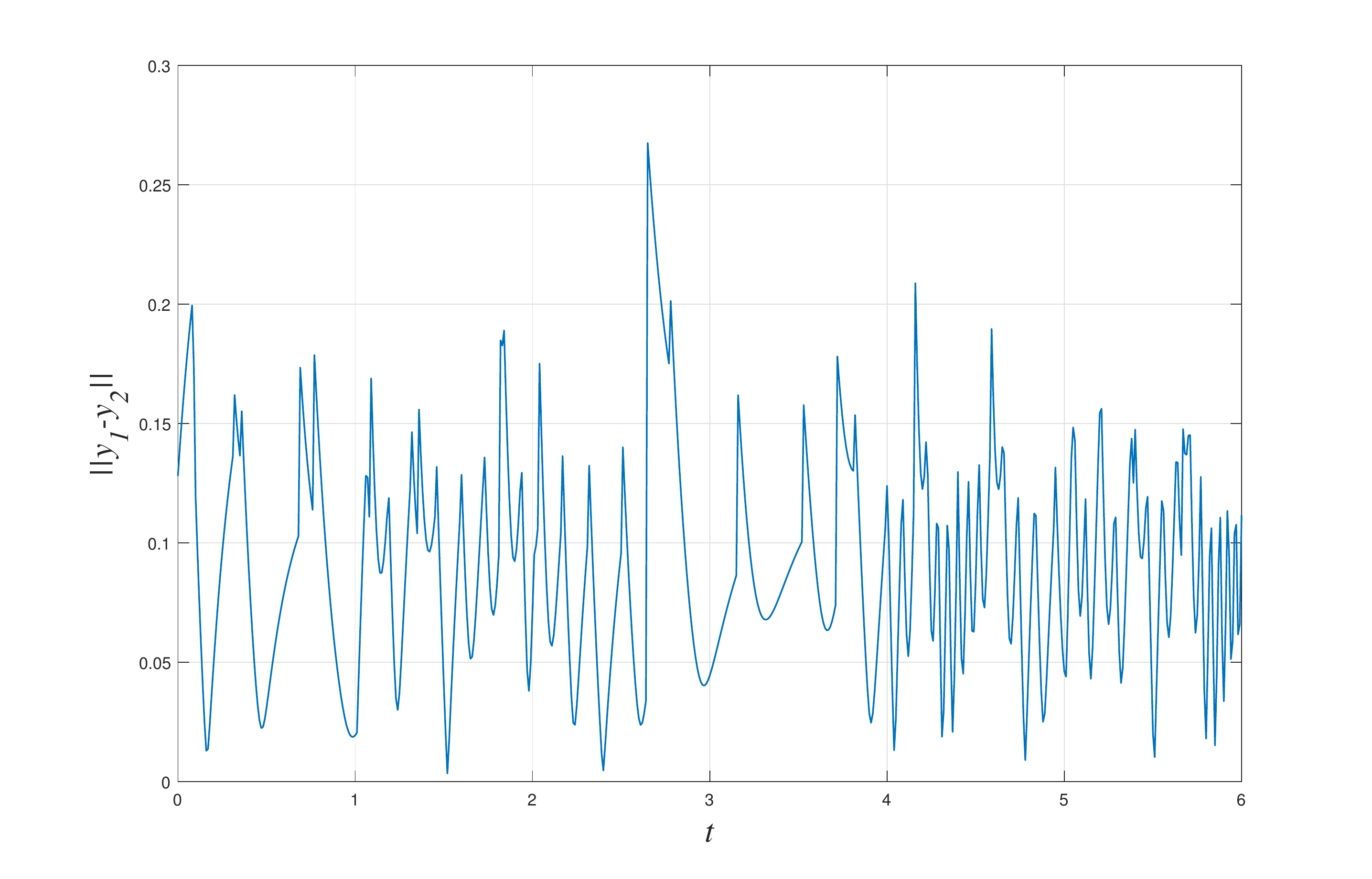}
\caption{The evolution of $\|y_1-y_2\|$.}\label{fig2}
\end{figure}

\begin{figure}[h!]
\centering
\includegraphics[height=4cm,width=9cm]{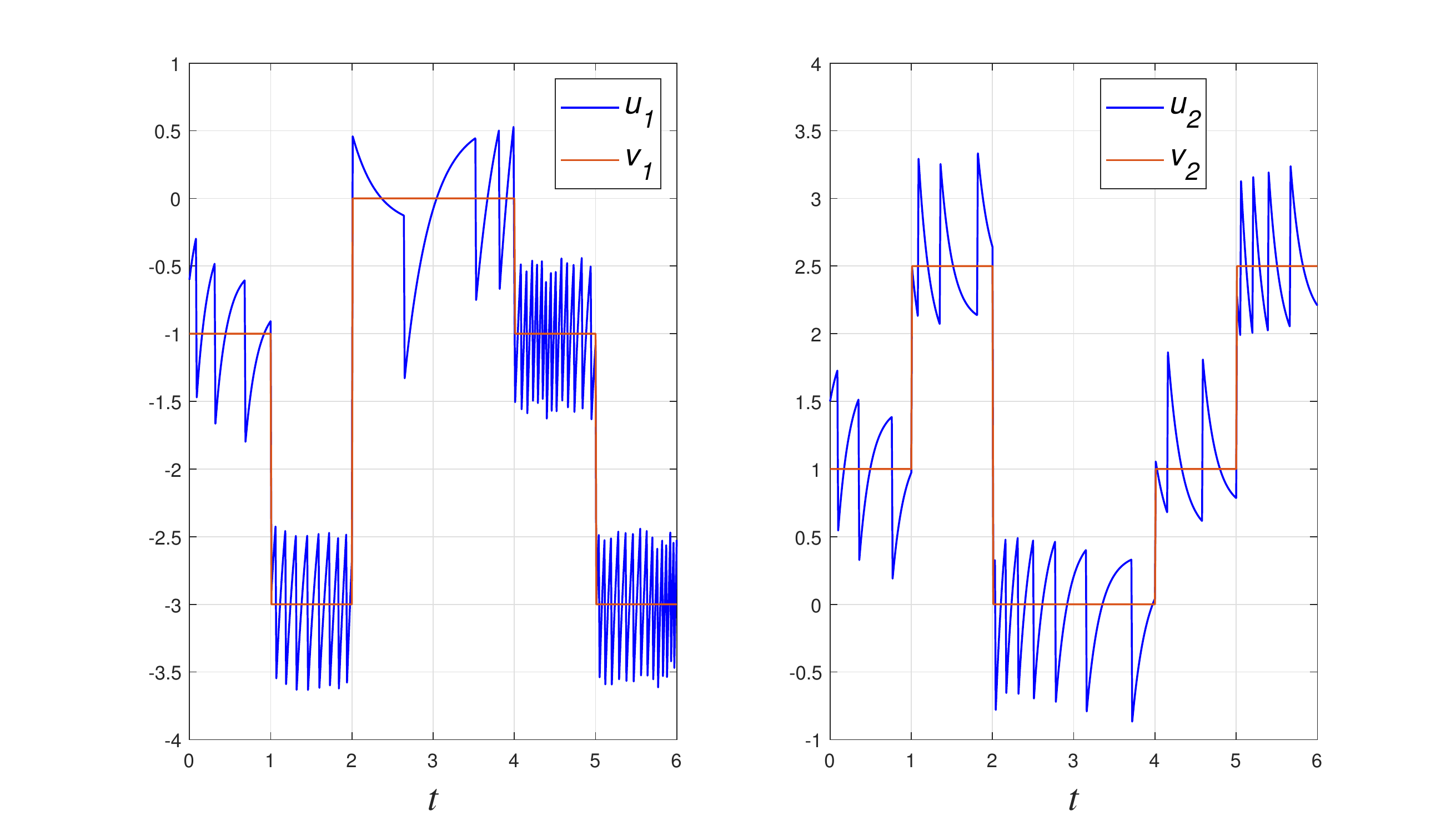}
\caption{The evolution of the inputs $u$ and $v$.}\label{fig3}
\end{figure}

The simulation results are shown in Figs. \ref{fig1}-\ref{fig3}. The trajectory $x_2$ of $\Sigma'$ is obtained by applying a piece-wise constant control input $v(t)$, and it is represented by the
solid red line in Fig. \ref{fig1} ($x_{2,1}, x_{2,2}$ are the two state components of $x_2$). The trajectory $x_1$ of $\Sigma$ is obtained via the control interface (\ref{uex2}), and it is represented by the solid blue line in Fig. \ref{fig1} ($x_{1,1}, x_{1,2}$ are the two state components of $x_1$). The evolution of the output error $\|y_1-y_2\|$ is depicted in Fig. \ref{fig2}, and one can see that the desired precision 0.5 is satisfied at all times. The evolution of the input components $v_1, v_2$ for the abstract system $\Sigma'$ and the input components $u_1, u_2$ for the concrete system $\Sigma$ is plotted in Fig. \ref{fig3}, respectively.

\section{Conclusion}

This paper involved the construction of discrete symbolic models for continuous-time nonlinear systems. Based on a new stability notion called controlled globally asymptotic/practical stability with respect to a set, it was shown that every continuous-time concrete system, under the condition that there exists an admissible control interface such that the augmented system can be made controlled globally practically stable with respect to a given set, is approximately simulated by its discrete (state-space) abstraction. In the future, the input quantization, external disturbances, and more efficient abstraction techniques, such as multi-scale abstraction will be taken into account.

\end{document}